\documentclass[prl,twocolumn,showpacs,floatfix,superscriptaddress]{revtex4-1}%
\usepackage{amsfonts,xcolor}
\usepackage{amsmath}
\usepackage{amssymb}
\usepackage{xcolor}

\usepackage{graphicx}
\usepackage{bm}
\usepackage{hyperref}
\usepackage[mathlines]{lineno}
\usepackage{float}
\graphicspath{{./figures/}}

\begin{document}


\title{Clustering and spatial distribution of mitochondria in dendritic trees}

\author{M. Hidalgo-Soria}
\affiliation{%
 Department of Physics, University of California, San Diego, La Jolla, CA 92093, USA
}%
\author{E. F. Koslover}
\email[]{ekoslover@ucsd.edu}
\affiliation{%
Department of Physics, University of California, San Diego, La Jolla, CA 92093, USA
}%

\date{\today}
\begin{abstract}
	
Neuronal dendrites form densely branched tree architectures through which mitochondria must be distributed to supply the cell's energetic needs. Dendritic  mitochondria circulate through the tree, undergoing fusion and fission to form clusters of varying sizes.  We present a mathematical model for the distribution of such actively-driven particles in a branched geometry. Our model demonstrates that `balanced' trees (wherein cross-sectional area is conserved across junctions and thicker branches support more bushy subtrees) enable symmetric yet distally enriched particle distributions and promote dispersion into smaller clusters. These results highlight the importance of tree architecture and radius-dependent fusion in governing the distribution of neuronal mitochondria.

\end{abstract}

\pacs{Valid PACS appear here}
\maketitle




Highly extended neuronal cells face the challenging task of spatially distributing organelles such as mitochondria, which provide a source of energy for neuronal function.  Defects in mitochondrial transport, localization, and interaction are associated with a number of neuropathologies, including  Alzheimer's and Parkinson's disease~\cite{van2009,schon2011mitochondria,mandal2019axonal}.
In neurons, mitochondria form a `social' network of variable-size clusters whose dynamic fusion and fission are thought to contribute to mitochondrial maintenance by helping deliver newly synthesized material from the soma throughout the cell~\cite{hoitzing2015,picard2021,agrawal2021}.
Mitochondrial structures range from highly interconnected architectures in yeast cells~\cite{viana2020}, to largely fragmented populations in axons~\cite{faitg20213d,pilling2006kinesin}, to networks on the border of percolation in many mammalian cell types~\cite{sukhorukov2012,zamponi2018,Holt2024}.
Dendritic mitochondrial networks consist of many disjoint clusters as well as small motile mitochondria~\cite{Donovan2022}.

The fusion of mitochondria into larger clusters is analogous to reversible aggregation and polymerization phenomena, previously  explored in the context of pathogenic protein aggregates~\cite{budrikis2014protein}, filaments~\cite{knowles2009}, and gels~\cite{ziff1980}.
  Such systems have traditionally been studied via kinetic mass-action models~\cite{blatz1945,Takayasu1988,Majumdar1998,Krapivsky1996,Krap2010}, which generally assume that  fusion, fission, absorption, and injection  of particles occur in an unstructured homogeneous space, and that the system can be treated as well-mixed.
Here we consider how the spatial architectures of dendritic arbors, as well as active transport and geometry-dependent fusion, modulate the formation of mitochondrial clusters.

\begin{figure}
\centering
   \includegraphics[width=0.4\textwidth]{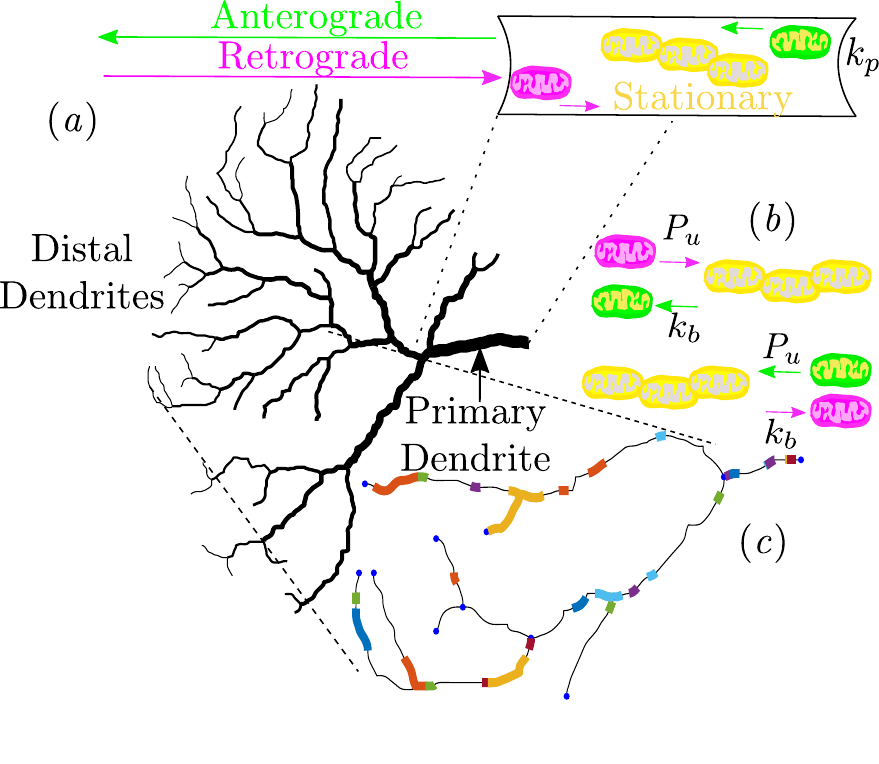}
   \caption{ 
     ({\em a}) Skeleton of a dendritic tree topology extracted from {\em Drosophila} HS neurons~\cite{Donovan2022}, with branch radii set to obey a balanced tree morphology. Inset: Schematic of anterograde (green), retrograde (magenta) and stationary (yellow) mitochondria  at the primary trunk. 
     Anterograde mitochondria are produced at rate $k_{p}$. 
     ({\em b}) Fusion and fission dynamics in the mitochondrial cluster model. Fusion occurs with probability $P_{u}$ during each passage event; fission at each cluster end occurs at constant rate $k_{b}$. ({\em c}) Simulation snapshot of mitochondrial clusters with different sizes on dendritic branches; different colors indicate distinct mitochondrial clusters. 
   }
  \label{fig:1}
\end{figure}

The dendritic arbor forms a bifurcating tree rooted at the cell body, with narrowing branches towards the distal tips (Fig.~\ref{fig:1}{\em a}). 
Live-cell imaging observations in {\em Drosophila} sensory HS neurons indicate both a stationary and a motile population of mitochondria that move processively in anterograde and retrograde directions~\cite{Donovan2022}. 
The steady-state distribution of mitochondria exhibits two key features: enrichment of mitochondrial volume density towards the distal tips, and symmetric volume densities between sister subtrees. The dendritic trees themselves approximately follow specific morphological scaling laws that allow the emergence of these distributions~\cite{Donovan2022}.
Namely, branch widths obey the Da Vinci law~\cite{richter1970} which preserves cross-sectional area across a junction ($r_0^2 = r_1^2 + r_2^2$, where $r_0$ is the radius of the parent branch and $r_1,r_2$ the radii of the two daughter branches). In addition, sister branch radii have areas in proportion to the `bushiness' ($B_i$, total branch length over depth) of the corresponding subtrees, according to $r_1^2/r_2^2 = B_1/B_2$. Trees that exhibit these two properties allow for symmetric volume densities of particles that undergo transport and radius-dependent halting~\cite{Donovan2022}; we shall refer to them hereafter as `balanced trees'. 

Here, we explore the formation of clusters due to fusion and fission of actively transported mitochondria on various tree morphologies. We demonstrate that balanced trees allow for a universal cluster distribution regardless of the tree branching pattern. Our results highlight the interplay between geometry and kinetics in governing the distribution of organelles in neuronal arbors.

Mitochondria are modeled as a population of discrete units that engage in transport, fission, and fusion (Fig.~\ref{fig:1}{\em b}). Single motile units move processively with velocity $\pm v$ (anterograde or retrograde) and reverse at branch tips. New mitochondria are produced at the soma (tree root) with rate $k_{p}$. Retrograde mitochondria that return to the root disappear. When a motile mitochondrion passes the tip of any mitochondrial cluster,
it has a probability of fusion: $P_{u,j}=A_{u}/r_{j}^{\gamma}$, with $r_{j}$ the radius of the  branch $j$, $\gamma$ a scaling exponent that governs sensitivity to branch width, and $A_{u}>0$. Fused clusters are always stationary.

Clusters can undergo fission with rate $k_b$ at each end, releasing a single motile mitochondrial unit while the remainder of the cluster remains stationary.
 A stationary single unit becomes motile again also with rate $k_{b}$ and is equally likely to go anterograde or retrograde. Anterograde mitochondria split at junctions in proportion to the cross-sectional area of the respective daughter branches, as observed in {\em Drosophila} sensory dendrites~\cite{Donovan2022}. 
 
 This model can be explicitly represented in agent-based simulations that track the motion of individual mitochondrial units and  the spatial extent of growing clusters (Fig.~\ref{fig:1}{\em c}; supplemental video SV1). 

To more efficiently explore the behavior of this dynamic system, we represent it in terms of a mass-action (mean-field) model, which treats all clusters as point-like particles and assumes that the cluster distributions can be described in terms of continuum concentration fields.
The mitochondria are represented by linear densities $\rho_{i,j}^\omega$ of  clusters with size $i$ located on branch $j$ in state $\omega = \left\{+,-,s\right\}$ (corresponding to anterograde, retrograde, and stationary particles, respectively). The time-evolution of these densities is given by the following set of equations, which incorporate all state transitions:
%
%
\begin{eqnarray}\label{eq:ME}
  \frac{\partial \rho_{1,j}^{s}}{\partial t} &=& -2k_{b}\rho_{1,j}^{s}+ 2k_{b}\rho_{2,j}^{s} -vP_{u,j}(\rho_{1,j}^{+}+\rho_{1,j}^{-})\rho_{1,j}^{s}, \nonumber \\
 \frac{\partial\rho_{1,j}^{\pm}}{\partial t} &=&  \mp v\frac{\partial \rho_{1,j}^{\pm}}{\partial x}-vP_{u,j}\rho_{1,j}^{\pm} \displaystyle \sum \limits _{i=1}^{\infty} \rho_{i,j}^{s} -2vP_{u,j}\rho_{1,j}^{-}\rho_{1,j}^{+}\nonumber\\
& + & k_{b}\rho_{2,j}^{s}+k_{b}\rho_{1,j}^{s}+k_{b}\displaystyle \sum \limits _{i=3}^{\infty} \rho_{i,j}^{s}, \\
 \frac{\partial\rho_{2,j}^{s}}{\partial t} &=& -vP_{u,j}(\rho_{1,j}^{+}+\rho_{1,j}^{-})\rho_{2,j}^{s}+2vP_{u,j}\rho_{1,j}^{-}\rho_{1,j}^{+}+2k_{b}\rho_{3,j}^{s} \nonumber\\
 & - & 2k_{b}\rho_{2,j}^{s}+vP_{u,j}(\rho_{1,j}^{+}+\rho_{1,j}^{-})\rho_{1,j}^{s}, \nonumber\\
 \frac{\partial\rho_{i,j}^{s}}{\partial t} &=& vP_{u,j}(\rho_{1,j}^{+}+\rho_{1,j}^{-})(\rho_{i-1,j}^{s}-\rho_{i,j}^{s})\nonumber \\
 &+&2k_{b}\rho_{i+1,j}^{s}-2k_{b}\rho_{i,j}^{s},\nonumber 
\end{eqnarray}

We define $\rho_{1,j}^{m} =\rho_{1,j}^{+} + \rho_{1,j}^{-}$ as the total linear density of motile mitochondria on branch $j$. Reflection boundary conditions at terminal nodes imply $\rho_{1,j}^{+} = \rho_{1,j}^{-}$ on all edges. In the parent trunk, the motile density is set by the boundary condition matching mitochondrial flux and production rate: $\rho_{1,0}^m = 2 k_p/v$. In daughter branches $k,l$ following the parent branch $j$, the motile linear density is found according to the  splitting rules: $\rho_{1,j}^m = \rho_{1,k}^m + \rho_{1,l}^m$ and $\rho_{1,k}^m/\rho_{1,l}^m = r_k^2/r_l^2$.

Putting together all the boundary conditions, the steady state solution of Eq.~\eqref{eq:ME}  follows (see details in SM):
\begin{eqnarray}\label{eq:csdbdv}
\rho_{i,j}^s &=& \frac{\rho_{1,j}^{m}}{2}(1+\alpha_{j} \rho_{1,j}^{m})\Big(\alpha_{j} \rho_{1,j}^{m}\Big)^{i-1}; \,\,\ i>1,
\end{eqnarray}
where $\alpha_{j}=vA_{u}/2k_{b}r_{j}^{\gamma}$.  The motile density in each branch is determined recursively from the parent branch according to the splitting rules at the junction. The branch widths play two separate roles in setting the mitochondrial densities: they determine the linear density of motile mitochondria in downstream branches, and they modulate the fusion probability through the scaling exponent $\gamma$. The volume density of clusters in each branch can be computed as $c_{i,j}^\omega = \rho_{i,j}^\omega/r_j^2$.

\begin{figure}
  \centering
  \includegraphics[scale=0.18]{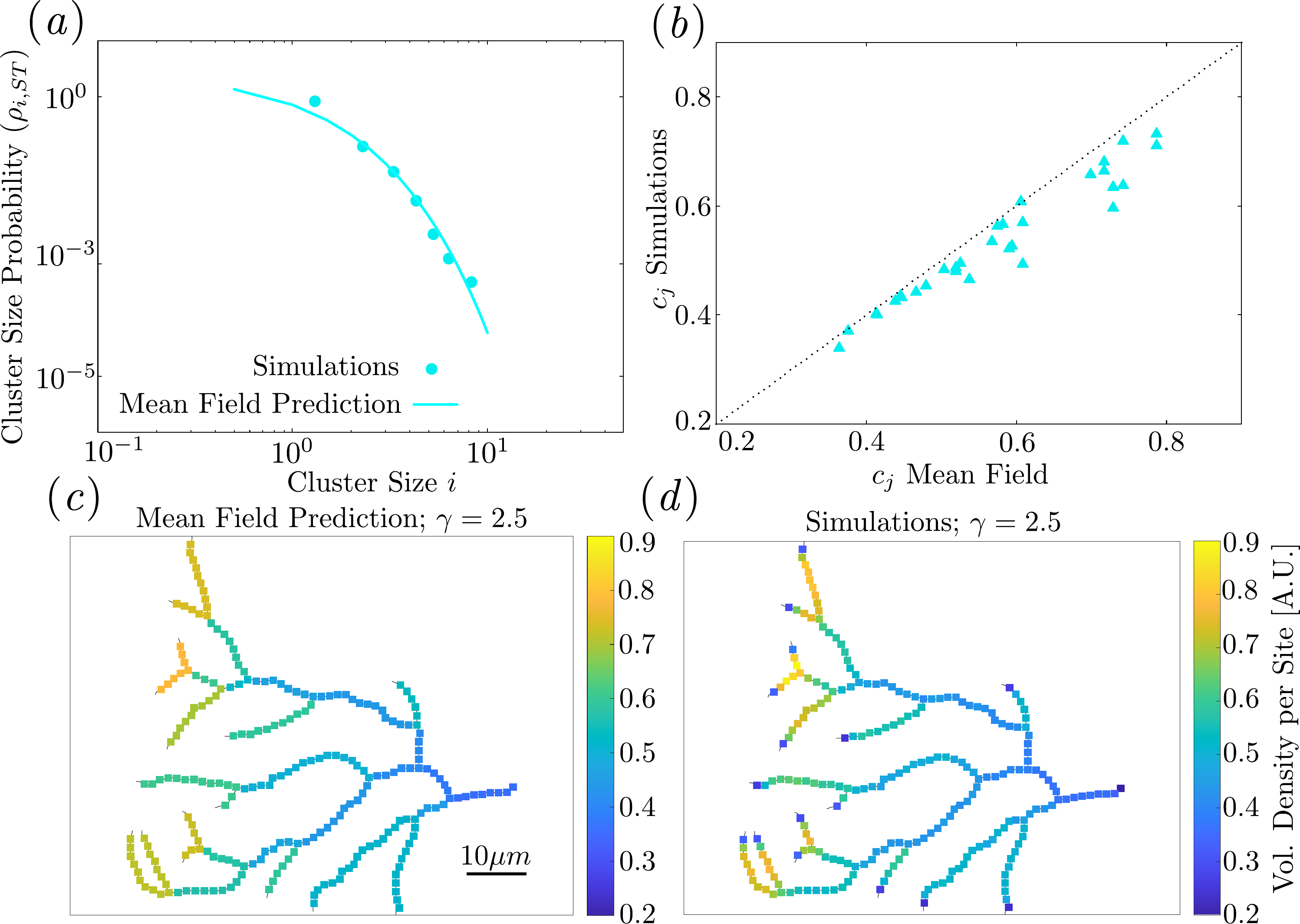}
	\caption{Mean-field model and simulations of cluster distributions on a balanced tree.
		 ({\em a}) Distribution of cluster sizes averaged across the tree.
		  ({\em b}) Average mitochondrial volume density is plotted for each branch using simulations (horizontal axis) and mean-field model (vertical axis). 
		  Dashed line indicates equality. ({\em c,d}) Spatial distribution of mitochondrial volume density $c_j = \sum_i i \rho_{i,j}/r_j^2$ from (\textit{c}) the mean-field model and ({\em d}) simulations. 
		  Parameters used throughout: $\gamma=2.5$, $k_{p}=0.19s^{-1}$, $k_{b}=0.01s^{-1}$, $v=0.45 \mu m/s$, $A_u = 0.01$, $r_\text{trunk} = 1$, unit size $1\mu$m in simulations. Average cluster size is $\left<i\right> = 1.3$ and average number of units is $M_T = 80$.
		   Results are averaged over 100 independent iterations.}
	\label{fig:2}
\end{figure}

In Fig.~\ref{fig:2}, we consider the cluster distribution on a smaller tree with branch radii set to obey a balanced tree morphology.
The overall cluster size distribution (integrated across the entire tree), exhibits an exponentially decaying tail, with a close match between simulations and the mean-field model (Fig.~\ref{fig:2}{\em a}). The higher mitochondrial density in distal branches (as expected for $\gamma>2$) is also predicted by the mean-field model (Fig.~\ref{fig:2}{\em c}).
Unlike the mean-field model, the simulations show a depletion of mitochondrial densities at the terminal tips of the branches (Fig.~\ref{fig:2}{\em d}). This effect arises from the finite size of mitochondrial units, which limits further fusion once a long cluster abuts the tip.
Nevertheless, the mean-field model encompasses the main features of the cluster size and spatial distributions in the simulated system.

We next determine how the total mitochondrial mass and its distribution into different-size clusters is governed by the interplay of kinetic parameters and the tree structure. To non-dimensionalize the model, we scale all branch widths relative to the trunk radius $r_\text{0}$. Branch lengths are scaled by the total depth of the tree $D_0$. For arbors where every tip has the same path length from the root, $D_0$ is equal to this path length. For more complicated arbor structures, the depth is defined recursively  according to $D_0 = \ell_0 + (L_1+L_2)/(L_1/D_1 + L_2/D_2)$, where $\ell_0$ is the length of the trunk, and $L_j, D_j$ are the total branch length and the depth of the daughter subtrees. For a balanced tree morphology, where sister branch radii are proportional to the subtree bushiness ($B_i = L_i/D_i$), the depth is directly related to the total tree volume, with $V_0 = D_0 r_0^2$ (see SM). 
Time units in the model are nondimensionalized by the time $D_0/v$ required for a mitochondrion to traverse the tree depth.

The resulting steady-state distribution of clusters on a fixed tree architecture is governed by two dimensionless parameters. The balance between new mitochondrial production and motile units traversing and exiting the tree is set by $\hat{k}_{p}=k_{p}D_{0}/v$. The balance between fusion and fission is set by $u=A_{u} v/(r^{\gamma}_{0} D_0 k_b)$. We note an analogy to prior models of binding and unbinding particles~\cite{Kegel2014}, with $\hat{k}_p$ serving as an effective fugacity (equivalent to the exponent of the chemical potential, driving more particles into the system) and $u$ serving as an effective association constant (balancing the preference for clustering versus fragmentation). 

It can be shown (see SM for details) that for any given tree structure, the linear densities of clusters ($m_{0,j}$) and of individual units ($m_{1,j}$) on each branch are given by: 
\begin{equation}
\begin{split}
m_{0,j} & = \frac{2 \xi_j \hat{k}_p}{D_0}\, g\left[u\hat{k}_p\xi_j\left(\frac{r_0}{r_j}\right)^\gamma\right], \\
m_{1,j} & = \frac{2 \xi_j \hat{k}_p}{D_0}\, f\left[u\hat{k}_p\xi_j\left(\frac{r_0}{r_j}\right)^\gamma\right], \\
g(x) & = 1+ \frac{1+x}{(1-x)}, \quad 
f(x) = 1+\frac{1+x}{(1-x)^2},
\end{split}
\end{equation}
where $\xi_j = \rho_{1,j}^m/\rho_{1,0}^m$ describes how motile mitochondria are diluted as they travel downstream, completely determined by the branch widths of the tree. If the tree structure is fixed, then the scaled total mitochondrial mass: $M_{T}/\hat{k}_p = \frac{1}{\hat{k}_p} \sum_j m_{1,j} \ell_j$ is a function of the reduced parameter $u\hat{k}_p$ and the fusion sensitivity exponent $\gamma$. Similarly, the average cluster size, $\left<i\right>_{T} = (\sum_j m_{1,j} \ell_j ) / (\sum_j m_{0,j} \ell_j )$ depends only on these two parameters and the tree structure.  

\begin{figure} 
  \centering
  \includegraphics[scale=0.18]{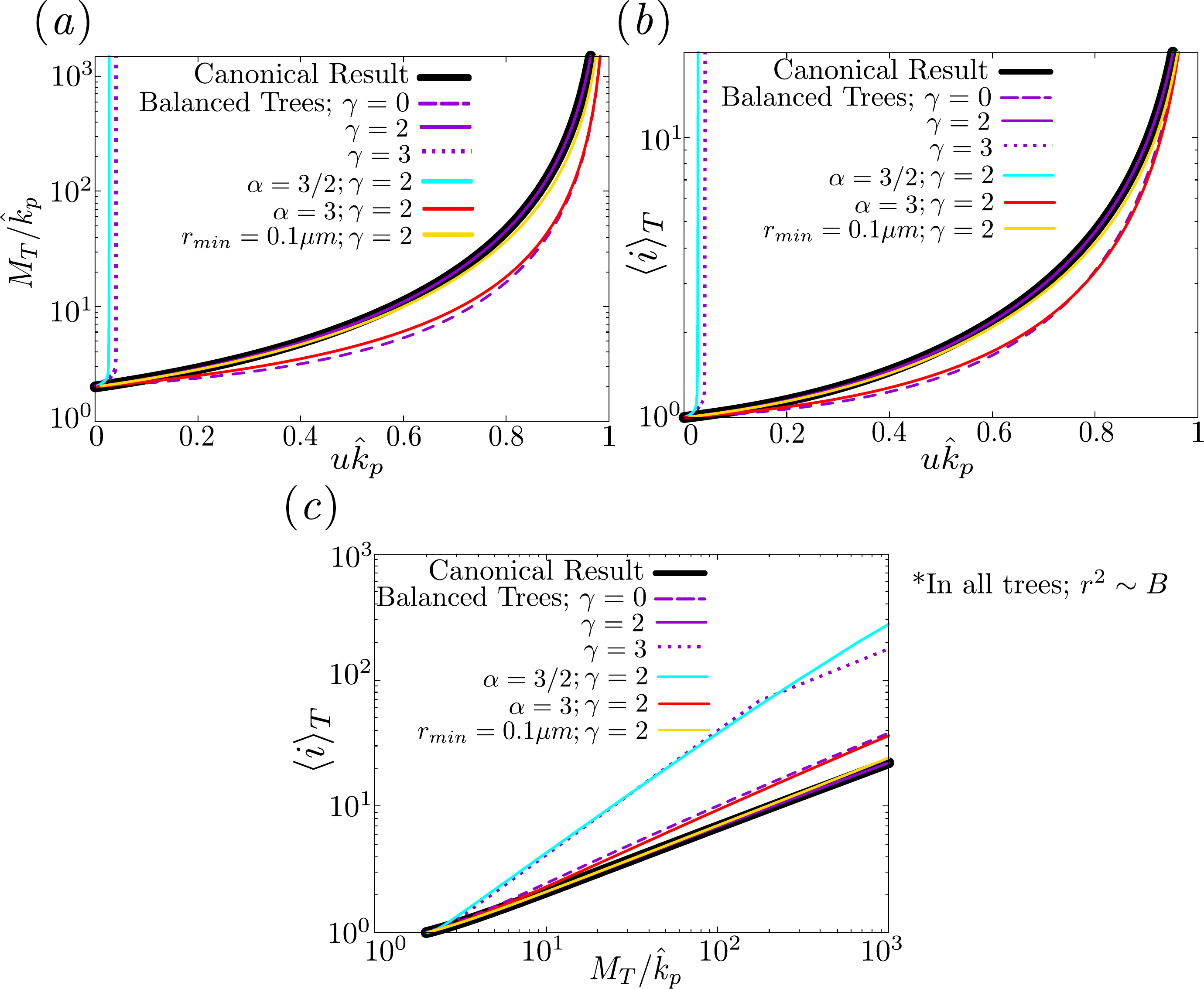}
  \caption[]{
    Mass-aggregation model results for the total mass and average cluster size with increasing fusion. ({\em a}) Scaled total mitochondrial mass plotted versus dimensionless parameter $u\hat{k}_p$.  ({\em b}) Average cluster size plotted versus  $u\hat{k}_p$. ({\em c}) Average cluster size plotted as a function of scaled mitochondrial mass. Black lines show canonical result on linear geometry with $\gamma=2$. All other curves are computed on an example HS dendritic arbor topology, with radii imposed according to specific scaling laws. 
    Purple lines show results for a balanced tree structure and different fusion scaling exponents: $\gamma=\lbrace 0,2,3\rbrace$ (dashed, solid, and dotted curves respectively). Cyan and red curves represent alternate radial scaling laws ($\alpha=3/2$ and $\alpha=3$, respectively), with $\gamma=2$. Yellow curves represent a balanced tree structure with an imposed minimum radius $r_\text{min}=0.1\mu$m.
} 
 \label{fig:MTACU}
\end{figure}

For trees that obey the Da Vinci Law, the splitting of motile mitochondria in proportion to branch area implies that $\xi_j = (r_j/r_0)^{2}$. Consequently, in the case of $\gamma=2$, the volume density of all cluster sizes is uniform across the tree.
The average cluster size is then given by a universal curve: $\left<i\right>_{T} = f(u\hat{k}_p)/g(u\hat{k}_p)$, regardless of the tree connectivity (see Supplemental Fig.~S1) 
The scaled total mitochondrial mass becomes $M_{T}/\hat{k}_p = 2f(u\hat{k}_p) V_0/(D_0 r_0^2)$, where $V_0$ is the total tree volume. For the special case of balanced trees, this scaled mass is also independent of the tree branching architecture. In particular, these universal expressions (black curves in Fig.~\ref{fig:MTACU}{\em a-b}) give the scaled mass and cluster size for particles on a simple linear domain. They are analogous to the canonical behavior found  in classic well-mixed mass aggregation models, which exhibit a phase transition with increasing fusion or production~\cite{budrikis2014protein,Takayasu1988}.

Both the total mass and cluster size diverge asymptotically when the parameter $u \hat{k}_p$ approaches a critical value, which for balanced trees is given by $(u\hat{k}_p)^* = (\text{min}\left\{r_j\right\}/r_0)^{\gamma-2}$. Thus, when fusion probabilities scale with cross-sectional area ($\gamma=2$), the critical value for diverging cluster size is $1$, whereas for steeper fusion sensitivity ($\gamma>2$), this critical value is lower (Fig.~\ref{fig:MTACU}{\em a-b}). 
We next consider the behavior of the total mass and average size for alternate tree structures that obey distinct radial scaling laws. For instance  when $\alpha= 3/2$ (Rall's Law~\cite{rall1959}), the total cross-sectional area is reduced across each junction. In such trees, the fusion probability increases more steeply towards the distal tips, inducing the formation of larger distal clusters. The critical value $(u \hat{k}_p)^*$ for divergent mass and cluster size is then substantially lower, so that comparatively low production rates can lead to substantial accumulation of distal mitochondria.
  When $\alpha=3$ (Murray's Law~\cite{murray1926}) the corresponding expansion of the cross sectional area at the distal zone reduces the fusion probability and leads to smaller clusters for the same production rate. 
  
In power-law models for tree structure, the cross-sectional area of the branches decreases exponentially from the primary trunk to the distal tips. However, there is likely a lower bound on the minimum radius of dendritic branches due to the mechanical limits of fitting at least one microtubule into each branch. Recent measurements in {\em Drosophila} class IV sensory neurons~\cite{liao2021} indicate that the branch radii can be approximated as obeying a modified Da Vinci relation: $r_1^2 + r_2^2 = r_0^2 + r_\text{min}^2$, with minimum radius $r_\text{min} \approx 0.1\mu$m. We consider the effect of incorporating this minimum radius into the model for mitochondrial cluster formation on the {\em Drosophila} HS dendritic arbors considered here, which have an average radius of $r_\text{trunk} \approx 3\mu$m in the primary trunk ~\cite{Donovan2022}. Imposing a balanced tree architecture yields an average radius of $\sim 0.3\mu$m in the distal branches. Consequently, the  minimum radius has only a small effect in widening distal tree branches and reducing cluster size (Fig.~\ref{fig:MTACU}).

 In general, cells can adjust rates of mitochondrial biogenesis to control their overall mitochondrial load~\cite{onyango2010regulation}. Thus, a fixed mitochondrial mass in the dendrite may be a more relevant control parameter than the production rate. 
 In the regime where clusters of substantial size are formed ($\left<i\right> \gtrsim 2$), a balanced tree architecture with fusion exponent $\gamma=2$ yields the smallest average cluster sizes (Fig.~\ref{fig:MTACU}{\em c}). A variety of HS dendritic trees with different branch width scaling laws, $\alpha=\lbrace 3/2,3\rbrace$, all give larger clusters for a given scaled mitochondrial mass (see Fig.~S1).
Thus, balanced trees with $\gamma \approx 2$ form the optimal structure to disperse the mitochondrial population into many small clusters, a direct consequence of the spatially uniform distributions that arise in such systems.

\begin{figure} 
  \centering
  \includegraphics[scale=0.18]{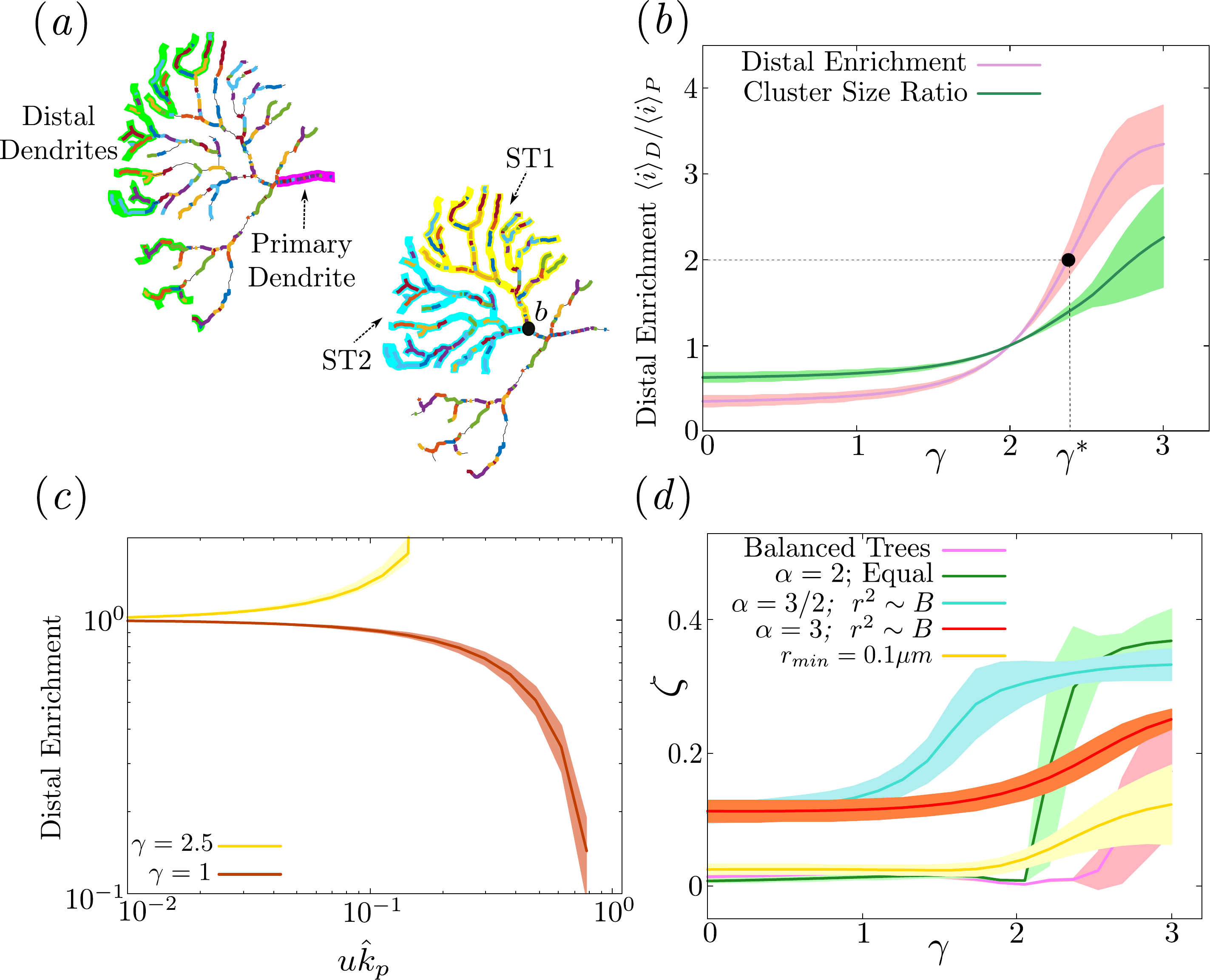}
    \caption[]{
      ({\em a}) Illustration of distal and proximal branches (left) and
      sister subtrees $ST_1$ (yellow), $ST_2$ (cyan) at junction $b$, in a sample HS dendritic tree. Colored strokes show mitochondrial cluster distribution from stochastic simulation snapshot.
      ({\em b}) Distal-to-proximal ratio in volume density (pink) and average cluster size (green) plotted against fusion scaling exponent $\gamma$. Average cluster size in whole tree is fixed to $\left<i\right> = 1.3$ by adjusting $A_u$.
       Experimentally observed distal enrichment (dashed line) corresponds to $\gamma^{\ast}=2.3$. 
      ({\em c}) Distal volume enrichment versus $u\hat{k}_{p}$ for $\gamma=\lbrace 1,2.5\rbrace$, with total number of mitochondrial units fixed to $M_{T}= 400$.
      ({\em d})  Root mean squared asymmetry ($\zeta$) of mitochondrial volume plotted against scaling exponent $\gamma$. Curves are shown for different radii scaling laws:  $\alpha=2, 3/2, 3$ (pink, cyan, orange) with sister radii split in proportion to subtree bushiness; $\alpha=2$ with equal sister radii (green); balanced trees with imposed minimum radius $r_{min}=0.1\mu m$ (yellow).  Average cluster size  is fixed to $\left<i\right> = 1.3$ by adjusting $A_u$.      
      All results are obtained from the mass-action model, averaged over $10$ tree topologies of distinct HS dendritic arbors~\cite{Donovan2022}; shadowed areas denote  standard deviation.
 } 
 \label{fig:DEAVMCTM1}
\end{figure}

The spatial accumulation of mitochondria in different regions of the tree can be quantified by
comparing clusters in distal branches (defined as those branches with path distance from the root greater than $70\%$ of the maximal value), versus those in the primary trunk (Fig.~\ref{fig:DEAVMCTM1}{\em a}). Focusing on balanced tree morphologies, the distal enrichment of both mitochondrial volume density and average cluster size are plotted in Fig.~\ref{fig:DEAVMCTM1}{\em b} as a function of the fusion scaling exponent $\gamma$. 
%



As previously noted, the scaling exponent $\gamma=2$ gives uniform volume densities throughout the tree. Higher values of $\gamma$ enrich the mitochondrial volume in distal branches while lower values lead to enrichment in the proximal trunk.
The ratio of distal to proximal cluster sizes also rises with increasing $\gamma$, albeit less steeply.
The 2-fold distal enrichment of volume density observed {\em in vivo}~\cite{Donovan2022} is obtained with fusion sensitivity exponent $\gamma^{\ast} \approx 2.3$. For this parameter, the average cluster size is expected to be a modest $38\%$ larger in distal versus proximal branches.


 We note that the probability of two axially passing particles coming in direct contact scales as $1/r^2$ when the size of the particles is small compared to the tube width, but should exhibit a steeper scaling in narrow tubes comparable to the particle size~\cite{mogre2020hitching}. Thus, values of $\gamma>2$ are to be expected for narrow distal branches.
Increasing the overall fusion probability (Fig.~\ref{fig:DEAVMCTM1}{\em c}) monotonically amplifies the enrichment of mitochondria in either the distal or proximal zones, depending on the value of $\gamma$.



Finally, we consider the predicted asymmetry of the mitochondrial distribution in sister subtrees. Measurements in HS neurons have demonstrated that the average mitochondrial volume density in sister subtrees tends to be equal, even when the subtree morphologies themselves are asymmetric~\cite{Donovan2022}. We define an asymmetry metric $\zeta=\sqrt{\frac{1}{N_{b}} \sum \limits _{b}(\frac{c_{b,ST1}-c_{b,ST2}}{c_{b,ST1}+c_{b,ST2}})^{2}}$, where $b$ is a junction site, $N_{b}$ is the total number of junctions in the tree, and $c_{b,ST1/2}$ are the average volume densities in the two subtrees emanating from the junction (Fig.~\ref{fig:DEAVMCTM1}{\em a}). For balanced trees, perfectly symmetric distributions are predicted for $\gamma=2$, with the asymmetry levels remaining quite low over a broad range of $\gamma$ values (Fig.~\ref{fig:DEAVMCTM1}{\em d}). Notably, other tree morphologies, such as a tree that obeys the Da Vinci Law but has equal radii between sister branches, or trees with alternate scaling exponents $\alpha$, 
 lead to much higher asymmetry when $\gamma>2$. The introduction of a minimum radius also increases the asymmetry of the mitochondrial distribution for high $\gamma$.
In all cases, reducing the system to the fragmented state ($u \hat{k}_p \rightarrow 0$ or $\left<i\right> = 1$) gives uniform volume densities.
 Overall, we show that a balanced tree morphology, approximately representative of HS dendrites, allows for a moderate amount of distal enrichment with a largely symmetric distribution of mitochondria.

Maintaining homeostasis of mitochondrial clusters in a neuronal arbor requires a balance of production, transport, fusion, and fission. The mass action model described here elucidates the key parameters setting the typical cluster size and overall mitochondrial accumulation: namely, the ratio of production and tree traversal rates ($\hat{k}_p$) and the ratio of fusion to fission ($u$).  The tree morphology affects mitochondrial distributions through both regulating the splitting of anterograde mitochondrial flux and modulating the probability of fusion between passing mitochondria (via the scaling exponent $\gamma$). Balanced trees architectures, which conserve branch area across junctions and split sister subtree trunk radii in proportion to bushiness, are shown to maintain symmetric distribution of mitochondria among subtrees, while allowing for enrichment at the distal tips. Such enrichment goes hand-in-hand with larger clusters in the distal region. Disperal into the smallest clusters is achieved on balanced trees, when fusion rates are inversely proportional to the cross-sectional area.


In this model system, mitochondrial distributions are shown to be highly sensitive to small changes in domain geometry or in the production and fusion kinetics. These results thus highlight key control parameters that can be tuned by the cell to alter the total volume, cluster architecture, and spatial distribution of the mitochondrial population that supplies the energetic needs of the dendritic arbor.

\paragraph*{\textbf{Acknowledgements}\\} 

We thank Erin Barnhart for helpful discussions. Support for this work was provided by the National Science Foundation, grants $\#2310229$ and $\#1848057$.
%

\bibliography{main}

\begin{thebibliography}{28}%
\makeatletter
\providecommand \@ifxundefined [1]{%
 \@ifx{#1\undefined}
}%
\providecommand \@ifnum [1]{%
 \ifnum #1\expandafter \@firstoftwo
 \else \expandafter \@secondoftwo
 \fi
}%
\providecommand \@ifx [1]{%
 \ifx #1\expandafter \@firstoftwo
 \else \expandafter \@secondoftwo
 \fi
}%
\providecommand \natexlab [1]{#1}%
\providecommand \enquote  [1]{``#1''}%
\providecommand \bibnamefont  [1]{#1}%
\providecommand \bibfnamefont [1]{#1}%
\providecommand \citenamefont [1]{#1}%
\providecommand \href@noop [0]{\@secondoftwo}%
\providecommand \href [0]{\begingroup \@sanitize@url \@href}%
\providecommand \@href[1]{\@@startlink{#1}\@@href}%
\providecommand \@@href[1]{\endgroup#1\@@endlink}%
\providecommand \@sanitize@url [0]{\catcode `\\12\catcode `\$12\catcode
  `\&12\catcode `\#12\catcode `\^12\catcode `\_12\catcode `\%12\relax}%
\providecommand \@@startlink[1]{}%
\providecommand \@@endlink[0]{}%
\providecommand \url  [0]{\begingroup\@sanitize@url \@url }%
\providecommand \@url [1]{\endgroup\@href {#1}{\urlprefix }}%
\providecommand \urlprefix  [0]{URL }%
\providecommand \Eprint [0]{\href }%
\providecommand \doibase [0]{http://dx.doi.org/}%
\providecommand \selectlanguage [0]{\@gobble}%
\providecommand \bibinfo  [0]{\@secondoftwo}%
\providecommand \bibfield  [0]{\@secondoftwo}%
\providecommand \translation [1]{[#1]}%
\providecommand \BibitemOpen [0]{}%
\providecommand \bibitemStop [0]{}%
\providecommand \bibitemNoStop [0]{.\EOS\space}%
\providecommand \EOS [0]{\spacefactor3000\relax}%
\providecommand \BibitemShut  [1]{\csname bibitem#1\endcsname}%
\let\auto@bib@innerbib\@empty
\bibitem [{\citenamefont {Van~Laar}\ and\ \citenamefont
  {Berman}(2009)}]{van2009}%
  \BibitemOpen
  \bibfield  {author} {\bibinfo {author} {\bibfnamefont {V.~S.}\ \bibnamefont
  {Van~Laar}}\ and\ \bibinfo {author} {\bibfnamefont {S.~B.}\ \bibnamefont
  {Berman}},\ }\href@noop {} {\bibfield  {journal} {\bibinfo  {journal} {Exp
  Neurol}\ }\textbf {\bibinfo {volume} {218}},\ \bibinfo {pages} {247}
  (\bibinfo {year} {2009})}\BibitemShut {NoStop}%
\bibitem [{\citenamefont {Schon}\ and\ \citenamefont
  {Przedborski}(2011)}]{schon2011mitochondria}%
  \BibitemOpen
  \bibfield  {author} {\bibinfo {author} {\bibfnamefont {E.~A.}\ \bibnamefont
  {Schon}}\ and\ \bibinfo {author} {\bibfnamefont {S.}~\bibnamefont
  {Przedborski}},\ }\href@noop {} {\bibfield  {journal} {\bibinfo  {journal}
  {Neuron}\ }\textbf {\bibinfo {volume} {70}},\ \bibinfo {pages} {1033}
  (\bibinfo {year} {2011})}\BibitemShut {NoStop}%
\bibitem [{\citenamefont {Mandal}\ and\ \citenamefont
  {Drerup}(2019)}]{mandal2019axonal}%
  \BibitemOpen
  \bibfield  {author} {\bibinfo {author} {\bibfnamefont {A.}~\bibnamefont
  {Mandal}}\ and\ \bibinfo {author} {\bibfnamefont {C.~M.}\ \bibnamefont
  {Drerup}},\ }\href@noop {} {\bibfield  {journal} {\bibinfo  {journal} {Front
  Cell Neurosci}\ }\textbf {\bibinfo {volume} {13}},\ \bibinfo {pages} {373}
  (\bibinfo {year} {2019})}\BibitemShut {NoStop}%
\bibitem [{\citenamefont {Hoitzing}\ \emph {et~al.}(2015)\citenamefont
  {Hoitzing}, \citenamefont {Johnston},\ and\ \citenamefont
  {Jones}}]{hoitzing2015}%
  \BibitemOpen
  \bibfield  {author} {\bibinfo {author} {\bibfnamefont {H.}~\bibnamefont
  {Hoitzing}}, \bibinfo {author} {\bibfnamefont {I.~G.}\ \bibnamefont
  {Johnston}}, \ and\ \bibinfo {author} {\bibfnamefont {N.~S.}\ \bibnamefont
  {Jones}},\ }\href@noop {} {\bibfield  {journal} {\bibinfo  {journal}
  {Bioessays}\ }\textbf {\bibinfo {volume} {37}},\ \bibinfo {pages} {687}
  (\bibinfo {year} {2015})}\BibitemShut {NoStop}%
\bibitem [{\citenamefont {Picard}\ and\ \citenamefont
  {Sandi}(2021)}]{picard2021}%
  \BibitemOpen
  \bibfield  {author} {\bibinfo {author} {\bibfnamefont {M.}~\bibnamefont
  {Picard}}\ and\ \bibinfo {author} {\bibfnamefont {C.}~\bibnamefont {Sandi}},\
  }\href@noop {} {\bibfield  {journal} {\bibinfo  {journal} {Neurosci Biobehav
  Rev}\ }\textbf {\bibinfo {volume} {120}},\ \bibinfo {pages} {595} (\bibinfo
  {year} {2021})}\BibitemShut {NoStop}%
\bibitem [{\citenamefont {Agrawal}\ and\ \citenamefont
  {Koslover}(2021)}]{agrawal2021}%
  \BibitemOpen
  \bibfield  {author} {\bibinfo {author} {\bibfnamefont {A.}~\bibnamefont
  {Agrawal}}\ and\ \bibinfo {author} {\bibfnamefont {E.~F.}\ \bibnamefont
  {Koslover}},\ }\href@noop {} {\bibfield  {journal} {\bibinfo  {journal}
  {{PLOS} Comput Biol}\ }\textbf {\bibinfo {volume} {17}},\ \bibinfo {pages}
  {e1009073} (\bibinfo {year} {2021})}\BibitemShut {NoStop}%
\bibitem [{\citenamefont {Viana}\ \emph {et~al.}(2020)\citenamefont {Viana},
  \citenamefont {Brown}, \citenamefont {Mueller}, \citenamefont {Goul},
  \citenamefont {Koslover},\ and\ \citenamefont {Rafelski}}]{viana2020}%
  \BibitemOpen
  \bibfield  {author} {\bibinfo {author} {\bibfnamefont {M.~P.}\ \bibnamefont
  {Viana}}, \bibinfo {author} {\bibfnamefont {A.~I.}\ \bibnamefont {Brown}},
  \bibinfo {author} {\bibfnamefont {I.~A.}\ \bibnamefont {Mueller}}, \bibinfo
  {author} {\bibfnamefont {C.}~\bibnamefont {Goul}}, \bibinfo {author}
  {\bibfnamefont {E.~F.}\ \bibnamefont {Koslover}}, \ and\ \bibinfo {author}
  {\bibfnamefont {S.~M.}\ \bibnamefont {Rafelski}},\ }\href@noop {} {\bibfield
  {journal} {\bibinfo  {journal} {Cell Syst}\ }\textbf {\bibinfo {volume}
  {10}},\ \bibinfo {pages} {287} (\bibinfo {year} {2020})}\BibitemShut
  {NoStop}%
\bibitem [{\citenamefont {Faitg}\ \emph {et~al.}(2021)\citenamefont {Faitg},
  \citenamefont {Lacefield}, \citenamefont {Davey}, \citenamefont {White},
  \citenamefont {Laws}, \citenamefont {Kosmidis}, \citenamefont {Reeve},
  \citenamefont {Kandel}, \citenamefont {Vincent},\ and\ \citenamefont
  {Picard}}]{faitg20213d}%
  \BibitemOpen
  \bibfield  {author} {\bibinfo {author} {\bibfnamefont {J.}~\bibnamefont
  {Faitg}}, \bibinfo {author} {\bibfnamefont {C.}~\bibnamefont {Lacefield}},
  \bibinfo {author} {\bibfnamefont {T.}~\bibnamefont {Davey}}, \bibinfo
  {author} {\bibfnamefont {K.}~\bibnamefont {White}}, \bibinfo {author}
  {\bibfnamefont {R.}~\bibnamefont {Laws}}, \bibinfo {author} {\bibfnamefont
  {S.}~\bibnamefont {Kosmidis}}, \bibinfo {author} {\bibfnamefont {A.~K.}\
  \bibnamefont {Reeve}}, \bibinfo {author} {\bibfnamefont {E.~R.}\ \bibnamefont
  {Kandel}}, \bibinfo {author} {\bibfnamefont {A.~E.}\ \bibnamefont {Vincent}},
  \ and\ \bibinfo {author} {\bibfnamefont {M.}~\bibnamefont {Picard}},\
  }\href@noop {} {\bibfield  {journal} {\bibinfo  {journal} {Cell Rep}\
  }\textbf {\bibinfo {volume} {36}} (\bibinfo {year} {2021})}\BibitemShut
  {NoStop}%
\bibitem [{\citenamefont {Pilling}\ \emph {et~al.}(2006)\citenamefont
  {Pilling}, \citenamefont {Horiuchi}, \citenamefont {Lively},\ and\
  \citenamefont {Saxton}}]{pilling2006kinesin}%
  \BibitemOpen
  \bibfield  {author} {\bibinfo {author} {\bibfnamefont {A.~D.}\ \bibnamefont
  {Pilling}}, \bibinfo {author} {\bibfnamefont {D.}~\bibnamefont {Horiuchi}},
  \bibinfo {author} {\bibfnamefont {C.~M.}\ \bibnamefont {Lively}}, \ and\
  \bibinfo {author} {\bibfnamefont {W.~M.}\ \bibnamefont {Saxton}},\
  }\href@noop {} {\bibfield  {journal} {\bibinfo  {journal} {Mol Biol Cell}\
  }\textbf {\bibinfo {volume} {17}},\ \bibinfo {pages} {2057} (\bibinfo {year}
  {2006})}\BibitemShut {NoStop}%
\bibitem [{\citenamefont {Sukhorukov}\ \emph {et~al.}(2012)\citenamefont
  {Sukhorukov}, \citenamefont {Dikov}, \citenamefont {Reichert},\ and\
  \citenamefont {Meyer-Hermann}}]{sukhorukov2012}%
  \BibitemOpen
  \bibfield  {author} {\bibinfo {author} {\bibfnamefont {V.~M.}\ \bibnamefont
  {Sukhorukov}}, \bibinfo {author} {\bibfnamefont {D.}~\bibnamefont {Dikov}},
  \bibinfo {author} {\bibfnamefont {A.~S.}\ \bibnamefont {Reichert}}, \ and\
  \bibinfo {author} {\bibfnamefont {M.}~\bibnamefont {Meyer-Hermann}},\
  }\href@noop {} {\bibfield  {journal} {\bibinfo  {journal} {{PLOS} Comput
  Biol}\ }\textbf {\bibinfo {volume} {8}},\ \bibinfo {pages} {e1002745}
  (\bibinfo {year} {2012})}\BibitemShut {NoStop}%
\bibitem [{\citenamefont {Zamponi}\ \emph {et~al.}(2018)\citenamefont
  {Zamponi}, \citenamefont {Zamponi}, \citenamefont {Cannas}, \citenamefont
  {Billoni}, \citenamefont {Helguera},\ and\ \citenamefont
  {Chialvo}}]{zamponi2018}%
  \BibitemOpen
  \bibfield  {author} {\bibinfo {author} {\bibfnamefont {N.}~\bibnamefont
  {Zamponi}}, \bibinfo {author} {\bibfnamefont {E.}~\bibnamefont {Zamponi}},
  \bibinfo {author} {\bibfnamefont {S.~A.}\ \bibnamefont {Cannas}}, \bibinfo
  {author} {\bibfnamefont {O.~V.}\ \bibnamefont {Billoni}}, \bibinfo {author}
  {\bibfnamefont {P.~R.}\ \bibnamefont {Helguera}}, \ and\ \bibinfo {author}
  {\bibfnamefont {D.~R.}\ \bibnamefont {Chialvo}},\ }\href@noop {} {\bibfield
  {journal} {\bibinfo  {journal} {Sci Rep}\ }\textbf {\bibinfo {volume} {8}},\
  \bibinfo {pages} {1} (\bibinfo {year} {2018})}\BibitemShut {NoStop}%
\bibitem [{\citenamefont {Holt}\ \emph {et~al.}(2024)\citenamefont {Holt},
  \citenamefont {Winter}, \citenamefont {Manley},\ and\ \citenamefont
  {Koslover}}]{Holt2024}%
  \BibitemOpen
  \bibfield  {author} {\bibinfo {author} {\bibfnamefont {K.}~\bibnamefont
  {Holt}}, \bibinfo {author} {\bibfnamefont {J.}~\bibnamefont {Winter}},
  \bibinfo {author} {\bibfnamefont {S.}~\bibnamefont {Manley}}, \ and\ \bibinfo
  {author} {\bibfnamefont {E.~F.}\ \bibnamefont {Koslover}},\ }\href {\doibase
  10.1101/2024.01.24.577101} {\bibfield  {journal} {\bibinfo  {journal}
  {bioRxiv}\ } (\bibinfo {year} {2024}),\
  10.1101/2024.01.24.577101}\BibitemShut {NoStop}%
\bibitem [{\citenamefont {Donovan}\ \emph {et~al.}(2022)\citenamefont
  {Donovan}, \citenamefont {Agrawal}, \citenamefont {Liberman}, \citenamefont
  {Kalai}, \citenamefont {Chua}, \citenamefont {Koslover},\ and\ \citenamefont
  {Barnhart}}]{Donovan2022}%
  \BibitemOpen
  \bibfield  {author} {\bibinfo {author} {\bibfnamefont {E.~J.}\ \bibnamefont
  {Donovan}}, \bibinfo {author} {\bibfnamefont {A.}~\bibnamefont {Agrawal}},
  \bibinfo {author} {\bibfnamefont {N.}~\bibnamefont {Liberman}}, \bibinfo
  {author} {\bibfnamefont {J.~I.}\ \bibnamefont {Kalai}}, \bibinfo {author}
  {\bibfnamefont {N.~J.}\ \bibnamefont {Chua}}, \bibinfo {author}
  {\bibfnamefont {E.~F.}\ \bibnamefont {Koslover}}, \ and\ \bibinfo {author}
  {\bibfnamefont {E.~L.}\ \bibnamefont {Barnhart}},\ }\href {\doibase
  10.1101/2022.07.01.497972} {\bibfield  {journal} {\bibinfo  {journal}
  {bioRxiv (in press at Cell Reports)}\ } (\bibinfo {year} {2022}),\
  10.1101/2022.07.01.497972}\BibitemShut {NoStop}%
\bibitem [{\citenamefont {Budrikis}\ \emph {et~al.}(2014)\citenamefont
  {Budrikis}, \citenamefont {Costantini}, \citenamefont {La~Porta},\ and\
  \citenamefont {Zapperi}}]{budrikis2014protein}%
  \BibitemOpen
  \bibfield  {author} {\bibinfo {author} {\bibfnamefont {Z.}~\bibnamefont
  {Budrikis}}, \bibinfo {author} {\bibfnamefont {G.}~\bibnamefont
  {Costantini}}, \bibinfo {author} {\bibfnamefont {C.~A.}\ \bibnamefont
  {La~Porta}}, \ and\ \bibinfo {author} {\bibfnamefont {S.}~\bibnamefont
  {Zapperi}},\ }\href@noop {} {\bibfield  {journal} {\bibinfo  {journal} {Nat
  Commun}\ }\textbf {\bibinfo {volume} {5}},\ \bibinfo {pages} {3620} (\bibinfo
  {year} {2014})}\BibitemShut {NoStop}%
\bibitem [{\citenamefont {Knowles}\ \emph {et~al.}(2009)\citenamefont
  {Knowles}, \citenamefont {Waudby}, \citenamefont {Devlin}, \citenamefont
  {Cohen}, \citenamefont {Aguzzi}, \citenamefont {Vendruscolo}, \citenamefont
  {Terentjev}, \citenamefont {Welland},\ and\ \citenamefont
  {Dobson}}]{knowles2009}%
  \BibitemOpen
  \bibfield  {author} {\bibinfo {author} {\bibfnamefont {T.~P.}\ \bibnamefont
  {Knowles}}, \bibinfo {author} {\bibfnamefont {C.~A.}\ \bibnamefont {Waudby}},
  \bibinfo {author} {\bibfnamefont {G.~L.}\ \bibnamefont {Devlin}}, \bibinfo
  {author} {\bibfnamefont {S.~I.}\ \bibnamefont {Cohen}}, \bibinfo {author}
  {\bibfnamefont {A.}~\bibnamefont {Aguzzi}}, \bibinfo {author} {\bibfnamefont
  {M.}~\bibnamefont {Vendruscolo}}, \bibinfo {author} {\bibfnamefont {E.~M.}\
  \bibnamefont {Terentjev}}, \bibinfo {author} {\bibfnamefont {M.~E.}\
  \bibnamefont {Welland}}, \ and\ \bibinfo {author} {\bibfnamefont {C.~M.}\
  \bibnamefont {Dobson}},\ }\href@noop {} {\bibfield  {journal} {\bibinfo
  {journal} {Science}\ }\textbf {\bibinfo {volume} {326}},\ \bibinfo {pages}
  {1533} (\bibinfo {year} {2009})}\BibitemShut {NoStop}%
\bibitem [{\citenamefont {Ziff}(1980)}]{ziff1980}%
  \BibitemOpen
  \bibfield  {author} {\bibinfo {author} {\bibfnamefont {R.~M.}\ \bibnamefont
  {Ziff}},\ }\href@noop {} {\bibfield  {journal} {\bibinfo  {journal} {J Stat
  Phys}\ }\textbf {\bibinfo {volume} {23}},\ \bibinfo {pages} {241} (\bibinfo
  {year} {1980})}\BibitemShut {NoStop}%
\bibitem [{\citenamefont {Blatz}\ and\ \citenamefont
  {Tobolsky}(1945)}]{blatz1945}%
  \BibitemOpen
  \bibfield  {author} {\bibinfo {author} {\bibfnamefont {P.}~\bibnamefont
  {Blatz}}\ and\ \bibinfo {author} {\bibfnamefont {A.}~\bibnamefont
  {Tobolsky}},\ }\href@noop {} {\bibfield  {journal} {\bibinfo  {journal} {J
  Phys Chem-us}\ }\textbf {\bibinfo {volume} {49}},\ \bibinfo {pages} {77}
  (\bibinfo {year} {1945})}\BibitemShut {NoStop}%
\bibitem [{\citenamefont {Takayasu}\ \emph {et~al.}(1988)\citenamefont
  {Takayasu}, \citenamefont {Nishikawa},\ and\ \citenamefont
  {Tasaki}}]{Takayasu1988}%
  \BibitemOpen
  \bibfield  {author} {\bibinfo {author} {\bibfnamefont {H.}~\bibnamefont
  {Takayasu}}, \bibinfo {author} {\bibfnamefont {I.}~\bibnamefont {Nishikawa}},
  \ and\ \bibinfo {author} {\bibfnamefont {H.}~\bibnamefont {Tasaki}},\ }\href
  {\doibase 10.1103/PhysRevA.37.3110} {\bibfield  {journal} {\bibinfo
  {journal} {Phys. Rev. A}\ }\textbf {\bibinfo {volume} {37}},\ \bibinfo
  {pages} {3110} (\bibinfo {year} {1988})}\BibitemShut {NoStop}%
\bibitem [{\citenamefont {Majumdar}\ \emph {et~al.}(1998)\citenamefont
  {Majumdar}, \citenamefont {Krishnamurthy},\ and\ \citenamefont
  {Barma}}]{Majumdar1998}%
  \BibitemOpen
  \bibfield  {author} {\bibinfo {author} {\bibfnamefont {S.~N.}\ \bibnamefont
  {Majumdar}}, \bibinfo {author} {\bibfnamefont {S.}~\bibnamefont
  {Krishnamurthy}}, \ and\ \bibinfo {author} {\bibfnamefont {M.}~\bibnamefont
  {Barma}},\ }\href {\doibase 10.1103/PhysRevLett.81.3691} {\bibfield
  {journal} {\bibinfo  {journal} {Phys. Rev. Lett.}\ }\textbf {\bibinfo
  {volume} {81}},\ \bibinfo {pages} {3691} (\bibinfo {year}
  {1998})}\BibitemShut {NoStop}%
\bibitem [{\citenamefont {Krapivsky}\ and\ \citenamefont
  {Ben-Naim}(1996)}]{Krapivsky1996}%
  \BibitemOpen
  \bibfield  {author} {\bibinfo {author} {\bibfnamefont {P.~L.}\ \bibnamefont
  {Krapivsky}}\ and\ \bibinfo {author} {\bibfnamefont {E.}~\bibnamefont
  {Ben-Naim}},\ }\href {\doibase 10.1103/PhysRevE.53.291} {\bibfield  {journal}
  {\bibinfo  {journal} {Phys. Rev. E}\ }\textbf {\bibinfo {volume} {53}},\
  \bibinfo {pages} {291} (\bibinfo {year} {1996})}\BibitemShut {NoStop}%
\bibitem [{\citenamefont {Krapivsky}\ \emph {et~al.}(2010)\citenamefont
  {Krapivsky}, \citenamefont {Redner},\ and\ \citenamefont
  {Ben-Naim}}]{Krap2010}%
  \BibitemOpen
  \bibfield  {author} {\bibinfo {author} {\bibfnamefont {P.~L.}\ \bibnamefont
  {Krapivsky}}, \bibinfo {author} {\bibfnamefont {S.}~\bibnamefont {Redner}}, \
  and\ \bibinfo {author} {\bibfnamefont {E.}~\bibnamefont {Ben-Naim}},\
  }\href@noop {} {\emph {\bibinfo {title} {A Kinetic View of Statistical
  Physics}}}\ (\bibinfo  {publisher} {Cambridge University Press},\ \bibinfo
  {year} {2010})\BibitemShut {NoStop}%
\bibitem [{\citenamefont {Richter}\ \emph {et~al.}(1970)\citenamefont {Richter}
  \emph {et~al.}}]{richter1970}%
  \BibitemOpen
  \bibfield  {author} {\bibinfo {author} {\bibfnamefont {J.~P.}\ \bibnamefont
  {Richter}} \emph {et~al.},\ }\href@noop {} {\emph {\bibinfo {title} {The
  notebooks of Leonardo da Vinci}}},\ Vol.~\bibinfo {volume} {2}\ (\bibinfo
  {publisher} {Courier Corporation},\ \bibinfo {year} {1970})\BibitemShut
  {NoStop}%
\bibitem [{\citenamefont {Weinert}\ \emph {et~al.}(2014)\citenamefont
  {Weinert}, \citenamefont {Brewster}, \citenamefont {Rydenfelt}, \citenamefont
  {Phillips},\ and\ \citenamefont {Kegel}}]{Kegel2014}%
  \BibitemOpen
  \bibfield  {author} {\bibinfo {author} {\bibfnamefont {F.~M.}\ \bibnamefont
  {Weinert}}, \bibinfo {author} {\bibfnamefont {R.~C.}\ \bibnamefont
  {Brewster}}, \bibinfo {author} {\bibfnamefont {M.}~\bibnamefont {Rydenfelt}},
  \bibinfo {author} {\bibfnamefont {R.}~\bibnamefont {Phillips}}, \ and\
  \bibinfo {author} {\bibfnamefont {W.~K.}\ \bibnamefont {Kegel}},\ }\href
  {\doibase 10.1103/PhysRevLett.113.258101} {\bibfield  {journal} {\bibinfo
  {journal} {Phys. Rev. Lett.}\ }\textbf {\bibinfo {volume} {113}},\ \bibinfo
  {pages} {258101} (\bibinfo {year} {2014})}\BibitemShut {NoStop}%
\bibitem [{\citenamefont {Rall}(1959)}]{rall1959}%
  \BibitemOpen
  \bibfield  {author} {\bibinfo {author} {\bibfnamefont {W.}~\bibnamefont
  {Rall}},\ }\href@noop {} {\bibfield  {journal} {\bibinfo  {journal} {Exp
  Neurol}\ }\textbf {\bibinfo {volume} {1}},\ \bibinfo {pages} {491} (\bibinfo
  {year} {1959})}\BibitemShut {NoStop}%
\bibitem [{\citenamefont {Murray}(1926)}]{murray1926}%
  \BibitemOpen
  \bibfield  {author} {\bibinfo {author} {\bibfnamefont {C.~D.}\ \bibnamefont
  {Murray}},\ }\href@noop {} {\bibfield  {journal} {\bibinfo  {journal} {J Gen
  Physiol}\ }\textbf {\bibinfo {volume} {9}},\ \bibinfo {pages} {835} (\bibinfo
  {year} {1926})}\BibitemShut {NoStop}%
\bibitem [{\citenamefont {Liao}\ \emph {et~al.}(2021)\citenamefont {Liao},
  \citenamefont {Liang},\ and\ \citenamefont {Howard}}]{liao2021}%
  \BibitemOpen
  \bibfield  {author} {\bibinfo {author} {\bibfnamefont {M.}~\bibnamefont
  {Liao}}, \bibinfo {author} {\bibfnamefont {X.}~\bibnamefont {Liang}}, \ and\
  \bibinfo {author} {\bibfnamefont {J.}~\bibnamefont {Howard}},\ }\href@noop {}
  {\bibfield  {journal} {\bibinfo  {journal} {P Natl Acad Sci}\ }\textbf
  {\bibinfo {volume} {118}},\ \bibinfo {pages} {e2022395118} (\bibinfo {year}
  {2021})}\BibitemShut {NoStop}%
\bibitem [{\citenamefont {Onyango}\ \emph {et~al.}(2010)\citenamefont
  {Onyango}, \citenamefont {Lu}, \citenamefont {Rodova}, \citenamefont {Lezi},
  \citenamefont {Crafter},\ and\ \citenamefont
  {Swerdlow}}]{onyango2010regulation}%
  \BibitemOpen
  \bibfield  {author} {\bibinfo {author} {\bibfnamefont {I.~G.}\ \bibnamefont
  {Onyango}}, \bibinfo {author} {\bibfnamefont {J.}~\bibnamefont {Lu}},
  \bibinfo {author} {\bibfnamefont {M.}~\bibnamefont {Rodova}}, \bibinfo
  {author} {\bibfnamefont {E.}~\bibnamefont {Lezi}}, \bibinfo {author}
  {\bibfnamefont {A.~B.}\ \bibnamefont {Crafter}}, \ and\ \bibinfo {author}
  {\bibfnamefont {R.~H.}\ \bibnamefont {Swerdlow}},\ }\href@noop {} {\bibfield
  {journal} {\bibinfo  {journal} {BBA-Mol Basis Dis}\ }\textbf {\bibinfo
  {volume} {1802}},\ \bibinfo {pages} {228} (\bibinfo {year}
  {2010})}\BibitemShut {NoStop}%
\bibitem [{\citenamefont {Mogre}\ \emph {et~al.}(2020)\citenamefont {Mogre},
  \citenamefont {Christensen}, \citenamefont {Niman}, \citenamefont
  {Reck-Peterson},\ and\ \citenamefont {Koslover}}]{mogre2020hitching}%
  \BibitemOpen
  \bibfield  {author} {\bibinfo {author} {\bibfnamefont {S.~S.}\ \bibnamefont
  {Mogre}}, \bibinfo {author} {\bibfnamefont {J.~R.}\ \bibnamefont
  {Christensen}}, \bibinfo {author} {\bibfnamefont {C.~S.}\ \bibnamefont
  {Niman}}, \bibinfo {author} {\bibfnamefont {S.~L.}\ \bibnamefont
  {Reck-Peterson}}, \ and\ \bibinfo {author} {\bibfnamefont {E.~F.}\
  \bibnamefont {Koslover}},\ }\href@noop {} {\bibfield  {journal} {\bibinfo
  {journal} {Biophys J}\ }\textbf {\bibinfo {volume} {118}},\ \bibinfo {pages}
  {1357} (\bibinfo {year} {2020})}\BibitemShut {NoStop}%
\end{thebibliography}%

\end{document}